\documentclass[onecolumn,prd,nofootinbib]{revtex4}
\usepackage{epsf}
\usepackage{dcolumn}
\usepackage{bm}
\usepackage{graphicx}

\usepackage[]{natbib}

\def\aj{AJ}%
\def\apj{ApJ}%
\def\apjl{ApJ}%
\def\apjs{ApJS}%
\def\aap{A\&A}%
\def\mnras{MNRAS}%
\def\nat{Nature}%
\def\gs{{_>\atop^{\sim}}}
\def\cgs{ ${\rm erg/cm}^{2}/{\rm s}$ } 

\newcommand{\be}{\begin{equation}}
\newcommand{\ee}{\end{equation}}

\def\4he{$^4$He}
\def\3he{$^3$He}
\def\7li{$^7$Li}

\def\ltsim{\raise 2pt \hbox {$<$} \kern-1.1em \lower 4pt \hbox {$\sim$}}
\def\gtsim{\raise 2pt \hbox {$>$} \kern-1.1em \lower 4pt \hbox {$\sim$}}

\begin{document}
\bibliographystyle{abbrvnat}

\title{Demography of high redshift AGN }

\author{Fabrizio Fiore$^1$, Simonetta Puccetti$^2$, Smita Mathur$^3$}
\affiliation{$^1$INAF - Osservatorio Astronomico di Roma} 
\affiliation{$^2$ASI SDC}
\affiliation{$^3$Ohio State University, USA}

\date{{\today}\\}

\begin{abstract}
High redshift AGN holds the key to understand the early structure
formation and to probe the Universe during its infancy.  We review the
latest searches for high-z AGN in the deepest X-ray field so far, the
Chandra Deep Field South (CDFS) 4 Msecond exposure. We do not confirm the
positive detection of a signal in the stacked Chandra images at the
position of z$\sim6$ galaxies recently reported by Treister and
collaborators \cite{treister:2011}. We present z$>3$ X-ray sources
number counts in the 0.5-2 keV band obtained joining CDFS faint
detections \cite{fiore:2011} with Chandra-COSMOS and XMM-COSMOS
detections. We use these number counts to make predictions for surveys
with three mission concepts: Athena, WFXT and a Super-Chandra.
\end{abstract}



\maketitle


\section{Introduction}

The study of high redshift AGN holds the key to understand the early
structure formation and probe the Universe during its infancy.  High-z
AGN can be used to investigate important issues such as: 1) the evolution of
the correlations between the black hole mass and the galaxy properties
(see e.g \cite{lamastra:2010} and references therein); 2) the AGN
contribution to the re-ionization, heating of the Inter-Galactic
Medium and its effect on structure formation
(e.g. \cite{boutsia:2011,mitra:2011} and references therein); 3)
scenarios for the formation of the black hole (BH) seeds, which will
eventually grow up to form the super-massive black holes (SMBHs) seen
in most galaxy bulges (e.g. \cite{volonteri:2005}; 4) investigate the
physics of accretion at high-z: is BH growth mainly due to relatively
few accretion episodes, as predicted in hierarchical scenarios (see
e.g. \cite{dotti:2010} and references therein), or by the so called
chaotic accretion (hundreds to thousands of small accretion episodes,
\cite{king:2008})?  5) BHs, being the structures with the fastest
(exponential) growth rate, can be used to constrain both the
expansion rate of the Universe and the growth rate of the primordial
perturbations at high-z, of competing cosmological scenarios
\cite{fiore:2010,lamastra:2011}.  6) The slope of the high-z AGN
luminosity function and of the SMBH mass function strongly depend on
the AGN duty cycle, and therefore their measure can constrain this
critical parameter. In turn, the AGN duty cycle holds information on
the AGN triggering mechanisms.  The evaluation of the evolution of the
AGN duty cycle can thus help in disentangling among competing
scenarios for AGN triggering and feeding \cite{fiore:2011}.

Large area optical and near infrared surveys such as the SDSS, the
CFHQS, the NOAO DWFS/DLS and the UKIDSS surveys have already been able
to discover large samples of z$>4.5$ QSOs
(e.g. \cite{richards:2006,glikman:2011}) and about 50 QSOs at $z>5.8$
(e.g. \cite{jiang:2009,willott:2010a,mortlock:2011}). The majority of
these high-z AGN are broad line, unobscured, high luminosity AGN. They
are likely the tips of the iceberg of the high-z AGN population.
Lower luminosity and/or moderately obscured AGN can, in principle, be
detected directly in current and future X-ray surveys.  Dedicated
searches for high-z AGN using both deep and wide area X-ray surveys
and a multi-band selection of suitable candidates can increase the
number of high-z AGN by a factor $>10$. In particular, it should be
possible to find hundreds rare high-z, high luminosity QSOs, in both
the all sky and deep eROSITA surveys (the 0.5-2 keV flux limit of the
all sky survey being the order of $10^{-14}$ \cgs, while that of the
deep survey, covering hundreds deg$^2$, should be 2-3 times deeper)
with a selection function much less biased than optical surveys.  To
constrain the faint end of the high-z AGN luminosity function, and
therefore the shape of the luminosity function and of the SMBH mass
function, we need to best exploit current and future deep surveys. The
Chandra Deep Field South is today the {\it premiere} field, with its 4
Msec and 3 Msec exposures obtained by Chandra and XMM respectively,
since 1999. Three different approaches have been so far applied to
this field: a) direct detection of sources in X-ray maps
(e.g. \cite{xue:2011}); b) search for X-ray emission at the position
of candidate high-z galaxies selected in the red and near infrared
bands \cite{fiore:2011}; c) stacking of X-ray counts at the position
of candidate high-z galaxies \cite{treister:2011}. Here we review all
three methods and give state of the art number counts of high-z AGN at
faint fluxes. We use these number counts to predict the number of
high-z AGN in possible future deep X-ray surveys.  A $H_0=70$ km
s$^{-1}$ Mpc$^{-1}$, $\Omega_M$=0.3, $\Omega_{\Lambda}=0.7$ cosmology
is adopted throughout.

\section{Stacking analysis of candidate high-z galaxies}

\begin{figure*}[t!]
\centering
\includegraphics[width=16cm]{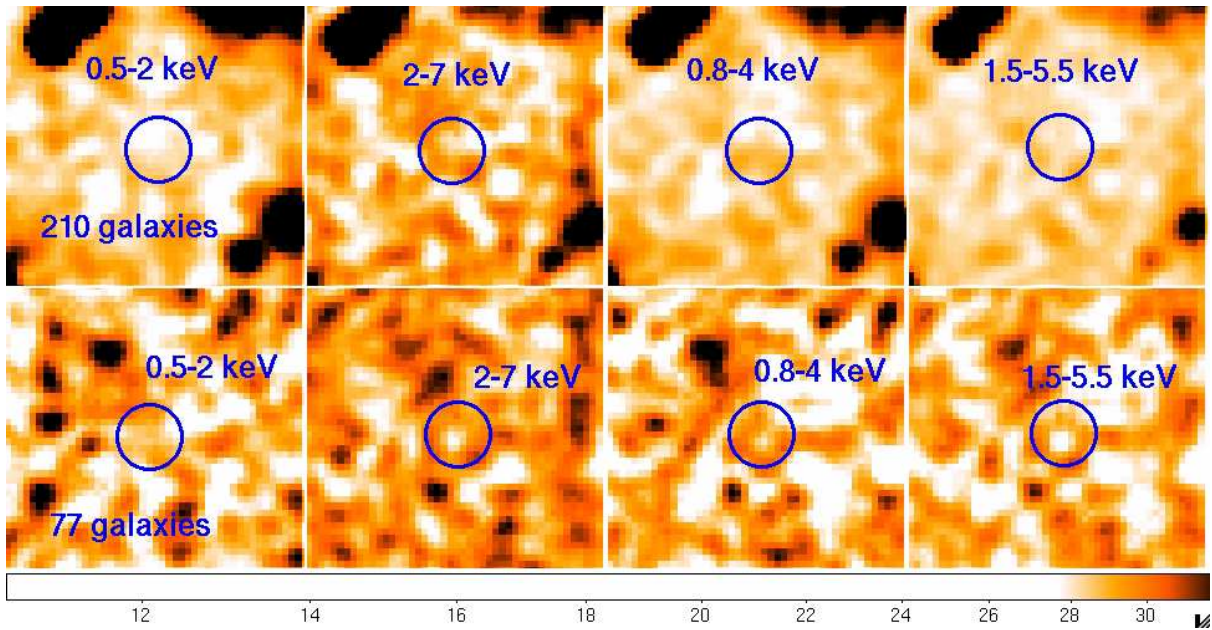}
\caption{Stacks of Chandra images at the position of 210 and 77
  Bouwens et al.  \cite{bouwens:2006} candidate z$\sim6-7$ galaxies in
  four energy bands: 0.5-2 keV, 2-7 keV, 0.8-4 keV and 1.5-5.5 keV.}
\label{imastacks}%
\end{figure*}

Recently Treister et al. \cite{treister:2011} published a positive
detection of X-ray counts in stacked Chandra images obtained adding
together the X-ray counts at the position of 197 candidate high-z
galaxies at z$\sim6$ in the CDFS and CDFN \cite{bouwens:2006}.  They
find 5$\sigma$ and 6.8$\sigma$ detections in the soft 0.5-2 keV and
hard 2-8 keV bands. Since the 2-8 keV flux they detect is about 9
times the 0.5-2 keV flux they infer that the majority of these faint
high-z galaxies host highly obscured, Compton thick AGN.  The total
rest frame 2-10 keV luminosity density implied by the Treister
\cite{treister:2011} result is $1.6\times10^{46}$ ergs/s/deg$^2$ at
z$\sim6$. In contrast, Fiore et al. \cite{fiore:2011} analyzed X-ray
counts at the position of the same Bouwens et al. \cite{bouwens:2006}
z$\sim6$ galaxies in the CDFS finding just one marginal
detection. Fiore et al.  \cite{fiore:2011} find that the z$\sim6$
luminosity function can be modeled using the standard double power law
shape:

\begin{equation}
\frac{{\rm d} \Phi (L_{\rm X})}{{\rm d Log} L_{\rm X}} 
= A [(L_{\rm X}/L_{*})^{\gamma 1} + (L_{\rm X}/L_{*})^{\gamma 2}]^{-1}.
\end{equation}

with $L^*=2\times 10^{44}$ ergs/s, $\gamma 1$=0.8 and $\gamma 2$=3.4
(the faint end slope is not truly constrained). By integrating this
luminosity function above a luminosity of $10^{42}$ erg/s one obtains
a total rest frame 2-10 keV luminosity density at z$\sim6$ of
$5.6\times10^{45}$ ergs/s/deg$^2$, a value $\sim2.8$ times smaller
than that reported by Treister et al. \cite{treister:2011}. We
investigated this discrepancy between the Treister \cite{treister:2011}
and Fiore \cite{fiore:2011} results. Once again we considered the
Bouwens et al.  \cite{bouwens:2006} sample of 371 candidate z$\sim6$
galaxies in the CDFS. Some of these galaxies happen to lie close to
bright X-ray sources identified with galaxies at a different redshift,
and must therefore be excluded from the analysis.  We considered two
exclusion radii, one similar to that used by Treister
\cite{treister:2011}, i.e. 22 arcsec, and another less conservative,
10 arcsec. In both cases we used the new Xue et al. \cite{xue:2011}
catalog of 740 directly detected X-ray sources.  We considered sources
at an offaxis angle $<8$arcmin, to avoid the inclusion of sources
observed with a too wide PSF.  We also considered only one galaxy when
we find 2 or more within 2 arcsec, not to count twice the contribution
from each single object. We finally excluded from the samples z$\sim6$
galaxies closer than 2 arcsec from lower redshift galaxies brighter
than zmag$=25$, which may contaminate the high redshift stacks.  The
final samples include 210 galaxies (10 arcsec exclusion radii) and 77
galaxies (22 arcsec exclusion radii). We performed stacks of Chandra
counts at the position of these galaxies in four energy bands: 0.5-2
keV, 2-7 keV, 0.8-4 keV and 1.5-5.5 keV.  The total exposure times for
the two sample are $\sim2.3\times 10^8$seconds (77 galaxies) and
$\sim6.3\times 10^8$seconds (210 galaxies).  Fig. \ref{imastacks} shows the
stacked images for the two samples in the four energy bands.  We do
not find a significant signal at the position of the galaxies in any
of these images.  Table 1 gives the PSF corrected 3$\sigma$ count rate
upper limits from the counts collected in boxes of 5 arcsec side (area
of 100 original pixels). As a comparison, Treister et
al. \cite{treister:2011} report a count rate $3.4\pm0.7\times 10^{-7}$
counts/s in the 0.5-2 keV band and $8.8\pm1.3\times 10^{-7}$ counts/s
in the 2-8 keV band.  Our more stringent upper limits are obtained for
the 210 galaxy sample in the 0.5-2 keV and 2-7 keV bands. These are
respectively comparable and 1.5 times lower than the Treister
\cite{treister:2011} claimed detections.

We can convert our count rate upper limits to a limit to the rest
frame 2-10 keV luminosity density following \cite{treister:2011}. We
find a 3$\sigma$ limit of $\sim10^{46}$ ergs/s/deg$^2$, lower than the
\cite{treister:2011} feature, but about twice the luminosity density
estimated by Fiore et al. \cite{fiore:2011}.

We recall that our analysis applies to the CDFS field alone, while the
\cite{treister:2011} result applies to the joined CDFS and CDFN
area. At least part of the discrepancy between Treister et
al. \cite{treister:2011} and our analysis could therefore be due to
cosmic variance. We also recall that for the sake of robustness our
stacking analysis is the simplest possible: counts at the position of
galaxies are added together, and aperture photometry is performed on
the stacked images without any optimization for off-axis dependent
PSF. Background is estimated in nearby regions, and, unlike
\cite{treister:2011}, no removal of positive fluctuations is
performed. While this simple technique does not probably push the
detection to the limit, it nevertheless produced valuable results in
the past when applied to samples of candidate, faint, 
Compton thick AGN \cite{fiore:2008,fiore:2009}.

\begin{table}
\caption{\bf 3$\sigma$ count rates upper limits}
\begin{tabular}{lcccc}
\hline
sample        &  0.5-2 keV & 2-7 keV & 0.8-4 keV & 1.5-5.5 keV \\
              & $10^{-7}$ cts/s & $10^{-7}$ cts/s & $10^{-7}$ cts/s & $10^{-7}$ cts/s\\
\hline
210 galaxies  & 3.4       & 5.8     & 6.2       & 6.2 \\
77 galaxies   & 5.9       & 9.7     & 9.6       & 10.1 \\
\hline
\end{tabular}

\end{table}

\section{High-z AGN number counts}

\begin{figure}[h]
\centering
\includegraphics[width=12.cm]{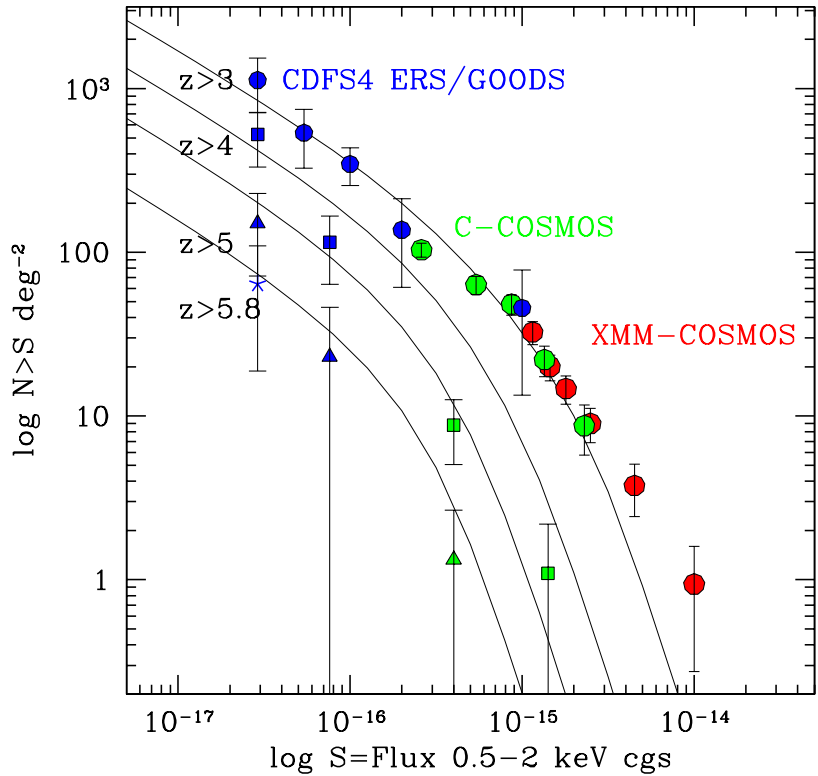}
\caption{Faint X-ray sources number counts in the 0.5-2 keV band. Blue
  points are from X-ray detections at the position of ERS and
  GOODS-MUSIC z$>3$ galaxies \cite{fiore:2011}; green points are from
  Chandra-COSMOS \cite{civano:2011}; red points are from XMM-COSMOS
  \cite{brusa:2009a}.  Circles = z$>3$ sources, squares = z$>4$
  sources, triangles = z$>5$ and star = z$>5.8$ sources.  The thin
  solid curves are model number counts based on the best fit high-z
  luminosity functions presented in \cite{fiore:2011}.}
\label{counts}%
\end{figure}

The analyses on the CDFS, CDFN, EGS and COSMOS fields provide samples
of individual sources detected, and therefore X-ray number counts of
faint high-z sources can be easily computed from these samples.
Fig. \ref{counts} show z$>3$ number counts from a compilation of
surveys: the Fiore et al. \cite{fiore:2011} survey of the ERS and
GOODS fields, and the Brusa et al. and Civano et al
\cite{brusa:2009a,civano:2011} XMM and Chandra surveys of the COSMOS
field. Black solid lines are model number counts obtained by
converting the \cite{fiore:2011} luminosity functions.  At the flux
limits reached by the deepest Chandra exposure (4 Mseconds) there are
$>1000$ z$>3$ AGN/deg$^2$, several hundreds of z$>4$ AGN/deg$^2$,
$>100$ z$>5$ AGN/deg$^2$, 20-100 z$>5.8$ AGN/deg$^2$ (the uncertainty
on the latter number is so large because it is based on just 2
candidate z$>$5.8 galaxies detected by Chandra in the small ERS
field). It is clear that to obtain a more robust demography of the
z$>$6 sky a search in a much wider area, such as the CANDELS area, is
mandatory, and requires spectroscopic confirmation of the X-ray
emitting, candidate z$>6$ galaxies. The CANDELS deep and wide surveys
cover a total of 130 arcmin$^2$ and 670 arcmin$^2$ to a depth of
H=27.8 and H$\sim26.5$ respectively. As a comparison, the ERS survey
covers an area of 50 arcmin$^2$ to a depth of H$\sim27$. The two
candidate z$>6$ ERS galaxies detected by Chandra in the ERS field are
faint, H=26.6 and H=27 sources. The GOODS source with z$>7$ in the Luo
et al. \cite{luo:2010} catalog has H=27.6.  The other z$>6$ ERS galaxy
with a marginal X-ray detection is brighter, H=23.8. In summary, we
expect 1-5 z$>6$ AGN in CANDELS deep and 4-20 z$>6$ AGN in CANDELS
wide. However, we note that a fraction of these sources will be at the
limit, or below, the H band sensitivity threshold of the wide
survey. As of today Chandra has spent of the order of 8 Mseconds on the
CANDELS fields, most of them on the CANDELS deep fields.  To reach the
sensitivity to detect the faint z$>6$ AGN in the wide area, additional
5-6 Mseconds are needed.  This is within reach of the Chandra observatory
in the next few years.  To make further progresses with Chandra,
i.e. quantitatively probe the first generation of accreting SMBH,
which would allow putting stringent constraints on SMBH formation
models \cite{madau:2001,lodato:2006,volonteri:2010b,begelman:2010},
and accretion scenarios
\cite{volonteri:2005,dotti:2010,fanidakis:2011,king:2008}, would
require at least triple the exposure times, i.e. 30-40 Mseconds on deep
surveys.  While this is certainly extremely expensive, it is not
technically unfeasible.

\section{Prediction for future surveys}

The Chandra limiting problem is that its sensitivity is very good on
axis, but degrades quickly at off-axis angles higher than a few
arcmin, making difficult and expensive in terms of exposure time to
cover with good sensitivity area larger than a few hundred 
arcmin$^2$.  A significant leap forward in the field would then be
obtained by an instrument capable of reaching the Chandra Msecond
exposure, on axis sensitivity (i.e flux limits in the range
$1-3\times10^{-17}$ \cgs) but on a factor of $>10$ wider field of view
(FOV). We consider here three possible mission concepts, making
predictions on the number of z$>4$, z$>5$ and z$>6$ faint x-ray
sources based on our best knowledge today, i.e. the number counts in
Fig. \ref{counts} and the luminosity functions in \cite{fiore:2011}.

1) Athena. This is a proposal for L class mission in the framework of
the ESA Cosmic Vision program. The baseline mission concept foresees
an effective area for imaging of the order of half square meter at 1-2
keV, a mirror PSF with half power diameter HPD$\sim10$ arcsec
(requirement, 5 arcsec goal), focal length 11m, FOV=0.17 deg$^2$
(25$\times$25 arcmin).  The observatory should be launched on a high
Earth orbit (HEO) or a L2 orbit, and therefore a rather high internal
backgrond is predicted (similar to the internal background measured by
the instruments on board XMM and Chandra, which are flying on HEO).

2) WFXT. The concept for a wide field X-ray telescope is quite old,
the first idea dating mid 90', and it evolved considerably over the
years.  We assumed a configuration similar to that in
\cite{murray:2009}, i.e an effective area of $>$half square meter at
1-2 keV, split in three mirror units, with HPD=10 arcsec
(requirement, goal 5 arcsec), and 5.5m focal length. Each mirror unit
is feeding a focal plane camera with FOV $\sim 1$ deg$^2$. We assumed
that the observatory is in a low Earth orbit (LEO), ensuring a low
internal background (similar to that of the instrument on board Swift
and Suzaku, which are flying on a LEO).

3) Super-Chandra.  This is a straw-man design for a mission with
imaging capabilities comparable to Chandra (i.e. arcsec HPD), but
using high throughput light weight mirrors (a concept pionereed by
Martin Elvis and Pepi Fabbiano some 15 years ago \cite{elvis:1996}).
Good imaging capabilities using thin glass or nickel shells may be
obtained by correcting the shell shape with actuators. Studies of
active X-ray mirrors have been performed in the past with good results
(see www.mssl.ucl.ac.uk/sxoptics). A SPIE conference had been devoted
to active X-ray mirrors in 2010 (Proceedings of SPIE 7803). Active
X-ray optics have been foreseen for extremely large throughput,
sub-arcsec future missions like Generetion-X
\cite{elvis:2006,odell:2010} or, more recently, for a square meter,
sub-arcsec mission (Vikhlinin et al. 2011, HEAD meeting).  Here we
assume a more modest throughput ($\sim3000$ cm$^2$ at 1-2 keV) and PSF
(1-2 arcsec HPD). We also assumed a limited FOV (0.1 deg$^2$) and a
LEO, which ensures a low internal background.

By using the above parameters we computed on axis sensitivies as a
function of the observing time (assuming a signal to noise ratio of 3
for source detection).  To make realistic predictions for the number of
high-z AGN expected, we assume that the effective area decreases
linearly from the center to the limit of the FOV by 50\%. 
The background includes particle induced internal background, as
measured on HEO and LEO, cosmic X-ray background (CXB), and low
temperature thermal X-ray background due to the local superbubble.
The internal background dominates over the X-ray background (CXB and
the local superbubble) above 0.5 keV in a HEO. Conversely, on LEO the
local superbubble dominates below 1 keV.  We finally assumed a total
net observing time of 12 Mseconds devoted to surveys, split in several
shorther observations, to cope with source confusion and optimize the
detection of z$>5$ sources with 2-10 keV luminosity $\gs10^{42}$
ergs/s.  The standard criterion for source confusion (40 beams per
source) translates in a flux limit for source confusion of
$\sim6\times10^{-17}$ \cgs in the 0.5-2 keV band for PSF HPD=10 arcsec
and just above $10^{-17}$ \cgs for HPD=5 arcsec. Source confusion is
not an issue for realistic exposure times for a PSF with HPD=2 arcsec or
below.

To estimate the faint X-ray sources number density we used the model
number counts in Fig \ref{counts}, based on the luminosity functions
presented in \cite{fiore:2011}. We conservatively assumed a faint-end
slope of the X-ray luminosity functions $\gamma 1$=0.6.  Table 2 gives
the predicted number of z=4--5, z=5--5.8 and z$>5.8$ sources, along
with their minimum 2-10 keV luminosity for several indicative mosaics
for the three mission concepts briefly described above.

\begin{table}
\caption{\bf Predicted number of faint, high-z X-ray sources}
\begin{tabular}{lclclll}
\hline
Mission concept & PSF HPD & Mosaics & total FOV    & z=4--5   & z=5--5.8  & z$>5.8$ \\
                & arcsec  &         & deg$^2$      & LX(z=5)  & LX(z=6)   &  LX(z=7) \\
\hline
Athena$^*$&  10  & 60x0.2Msec  & 10   & 940  $>$43.3  & 480  $>$43.5 & 250  $>$43.6  \\
Athena$^*$&  5   & 6x2Msec     & 1.0  & 360  $>$42.5  & 210  $>$42.6 & 125  $>$42.8  \\
Athena    &  5   & 40x0.3Msec  & 7.0  & 1100 $>$43    & 650  $>$43.1 & 360  $>$43.2  \\
\hline                                       
WFXT$^*$  &  10  & 24x0.5Msec  & 24   & 2300 $>$43.2  & 1300 $>$43.4 & 600  $>$43.5  \\
WFXT$^*$  &  5   & 4x3Msec     & 4    & 1200 $>$42.5  & 700  $>$42.6 & 400  $>$42.8 \\
WFXT      &  5   & 60x0.2Msec  & 60   & 6000 $>$43.35 & 3200 $>$43.4 & 1600 $>$43.5 \\
\hline                                       
S-Chandra &  2   & 6x2Msec     & 0.6  & 310  $>$42.2  & 185  $>$42.3 & 110  $>$42.5   \\
S-Chandra &  2   & 24x0.5Msec  & 2.4  & 390  $>$43    & 220  $>$43.1 & 125  $>$43.2  \\
S-Chandra &  1   & 2x6Msec     & 0.2  & 175  $>$41.8  & 100  $>$42.0 & 65   $>$42     \\
S-Chandra &  1   & 6x2Msec     & 0.6  & 350  $>$42.1  & 210  $>$42.2 & 120  $>$42.4   \\
\hline
\end{tabular}

$^*$ close to confusion limit (40 beams per source).
\end{table}

It must be noted that the uncertainties on the number of sources in
Table 2 is large. It is at least a factor of two at z=4-6 and even
bigger at z$>6$ (factor of 3 lower limit and a factor of 2 upper
limit).  The obvious message of Table 2 is that a wide field greatly
helps in searching for high-z AGN. This is probably the only solution
to collect samples of thousands X-ray AGN at z$>4$. However, even a
PSF as good as 5 arcsec HPD does not allow searching for sources
fainter than $10^{43}$ ergs/s at z$>5-7$. This means that only a
mission with Chandra-like PSF but much higher throughput ($>5\times$
Chandra effective area at 1-2 keV) would be able to target normal
starforming galaxies and mini-quasars at z=6-7.  On one hand, a 2-10
keV luminosity of $10^{42}$ ergs/s at z=7, reachable by extradeep
exposures with a 1 arcsec PSF Super-Chandra, would be produced by a
$7\times10^{5}$ SMBH emitting at its Eddington luminosity (assuming a
bolometric correction of 10). Even smaller masses may be probed, if
the accretion is super-Eddington. A Super-Chandra would then be able
to directly search for the first generation of SMBH produced by
monolithic collapse of $\gs10^5$ M$\odot$ gas clouds to BH
\cite{lodato:2006,lodato:2007,begelman:2010,volonteri:2010b}.  On the
other hand, L(2-10)=$10^{42}$ ergs/s is also produced by galaxies
which form stars at a rate of $\gs200$M$_\odot$/yr. Since at such high
redshifts X-ray emission should mainly be due to high mass X-ray
binaries, X-ray high-z galaxies could then be used to constrain the
initial mass function at the epoch of galaxy formation.  A
Super-Chandra would then be able to open two brand-new fields in
structure formation. Of course it is not casual that the considered
configuration for a Super-Chandra is able to reach these goals. Going
back from scientific requirements to mission parameters, the goal of
detecting sources with a 2-10 keV luminosity of $\sim10^{42}$ at
z$\sim7$, in feasible exposure times, requires an effective area
$\sim3000$ cm$^2$, given a PSF HPD$\sim1$ arcsec and assuming a LEO
low internal background.

Unfortunately a Super-Chandra is beyond the horizon of the present
decade, both because technological and programmatic
issues. Furtheremore, it is not clear whether a WFXT is truly feasible
with such huge 1 deg$^2$ FOV and large throughput, and in any case it
does not appear to be a priority in the latest US Decadal Survey
(http://sites.nationalacademies.org/bpa/BPA\_049810) nor in the ESA
Cosmic Vision program.  Conversely, Athena is a study mission for an L
class mission in the framework of the ESA Cosmic Vision program. A
decision for CV L class mission should be taken in February 2012. If
positive, Athena could be implemented for the first years of the next
decade. Although not reaching exquisite, Chandra-like image quality,
nor extra-large field of view, Athena would be able to give a
substantial contribution on the knowledge of the high-z Universe, with
hundreds to a thousand z$>4$ X-ray AGN (an improvement by a factor
10-100 with respect to today situation) and tens to hundreds z$>5.8$
X-ray selected AGN (today there are only 3-4 candidate z$>6$ X-ray
selected AGN in the litterature
\cite{luo:2010,salvato:2011,fiore:2011}).

\acknowledgments 

This work was supported by ASI/INAF contracts I/024/05/0 and
I/009/10/0. This work is based on observations made with NASA X-ray
observatory Chandra. We thank the Chandra Director's office for
allocating the time for these observations. X-ray data were obtained
from the archive of the Chandra X-ray Observatory Center, which is
operated by the Smithsonian Astrophysical Observatory.


\end{document}